\documentclass[multphys,vecphys]{svmult}

\usepackage{makeidx}         % allows index generation
\usepackage{graphicx}        % standard LaTeX graphics tool
\usepackage{multicol}        % used for the two-column index
\usepackage[bottom]{footmisc}% places footnotes at page bottom
\usepackage{url}
\makeindex             % used for the subject index
\begin{document}

\title*{Analytical and numerical studies of fluid instabilities in
relativistic jets.}  
\titlerunning{Fluid instabilities in relativistic
jets.}
\author{M. Perucho\inst{1}, A.P. Lobanov\inst{1} \and
J.M.$^{\underline{a}}$Mart\'{\i}\inst{2}}

\institute{Max-Planck-Institut f\"ur Radioastronomie. Auf dem
H\"ugel 69 53121 Bonn, Germany. \texttt{perucho@mpifr-bonn.mpg.de}
 \texttt{alobanov@mpifr-bonn.mpg.de}
\and Departament d'Astronomia i Astrof\'{\i}sica. Universitat de
Val\`encia. C/ Dr. Moliner 50 46100 Burjassot, Val\`encia, Spain.
\texttt{jose-maria.marti@uv.es}}
\maketitle

\begin{abstract}
Relativistic outflows represent one of the best-suited tools to probe
the physics of AGN.  Numerical modelling of internal structure of the
relativistic outflows on parsec scales provides important clues about
the conditions and dynamics of the material in the immediate vicinity
of the central black holes in AGN. We investigate possible causes of
the structural patterns and regularities observed in the parsec-scale
jet of the well known quasar 3C\,273. We compare the model with the
radio structure observed in 3C\,273 on parsec scales using very long
baseline interferometry (VLBI) and constrain the basic properties of the
flow. Our results show that Kelvin-Helmholtz instabilities are the
most plausible mechanism to generate the observed structures.
\end{abstract}

%Your text goes here. Separate text sections with the standard \LaTeX\
%sectioning commands.

\section{Introduction}
\label{sec:1} 3C\,273 is the brightest quasar known. Its
relative proximity ($z=0.158$, \cite{str92}) has made it a
paradigmatic object studied throughout the entire spectral range
\cite{cou98}.  One of the most impressive features of 3C\,273 is an
apparently one-sided relativistic outflow emanating from the quasar
nucleus \cite{cou98} and extending up to a deprojected distance of
about 60\,kpc.  The jet in 3C\,273 has been observed on parsec scales
using ground and space VLBI \cite{kri00,abr06,lz01,asa02}. Space VLBI
observations made at 5\,GHz with the VSOP\footnote{VLBI Space
Observatory Program} revealed a double helical patterns inside the
parsec-scale jet in 3C\,273 \cite{lz01}. Linear perturbation analysis
of Kelvin-Helmholtz (K-H) instability \cite{har00} applied to model
these structures yielded an accurate estimate of the bulk parameters
of the flow \cite{lz01}. A different interpretation suggests that a
helical magnetic field could generate such a structure
\cite{asa02}. We study the first possibility. A bulk Lorentz factor
$W=2.1$ obtained in \cite{lz01} is below the values inferred from
superluminal motions observed in the jet ($W=5-10$, \cite{abr06}),
which suggests that the K-H instability develops in a slower
underlying flow, while the fast components are shock waves inside the
jet. Four K-H modes were found acting on parsec scales in the 
jet in 3C\,273: the second helical body mode with a wavelength of 2
mas, the first elliptical and helical body modes, both with
wavelengths close to 4 mas, and the elliptical surface mode, with a
wavelength of 12 mas. Additionally a structure with a wavelength of 18\,mas
was interpreted as an externally driven helical surface mode.
Our previous
works \cite{pe04a,pe04b} have shown that numerical simulations can be
used effectively to connect the linear perturbation analysis with studies
of non-linear regime of K-H instability.
Our aim here is to use the jet parameters determined in \cite{lz01} as
initial conditions for numerical RHD simulations of a perturbed jet and
compare the numerical results with the observed structures.  
% Always give a unique label
% and use \ref{<label>} for cross-references
% and \cite{<label>} for bibliographic references
% use \sectionmark{}
% to alter or adjust the section heading in the running head

\section{Numerical simulations}

\subsection{Simulation 3C\,273-A}

We start with a steady jet with a
Lorentz factor $W=2.1$, a density contrast with the external medium
$\eta=0.023$, a sound speed $c_{s,j}=0.53\, c$ in the jet and
$c_{s,a}=0.08\, c$ in the external medium, and the perfect gas
equation of state (with adiabatic exponent $\gamma=4/3$). Assuming an
angle to the line of sight $\theta=15^\circ$,
%and redshift $z=0.158$
%($1\,\rm{mas}=2.43\,\rm{pc}$), 
the observed jet extends up to $\approx 170\,\rm{pc}$.  With the jet
radius $R_\mathrm{j} = 0.8$\,pc \cite{lz01}, the numerical grid covers
$211\, R_\mathrm{j}$ (axial) $\times$ $8\,R_\mathrm{j}$ $\times$
$8\,R_\mathrm{j}$ (transversal), i.e., $169\,\rm{pc}\times
6.4\,\rm{pc}\times 6.4\,\rm{pc}$. The resolution is $16\, {\rm
cells}/R_\mathrm{j}$ in the transversal direction and $4\,{\rm
cells}/R_\mathrm{j}$ in the direction of the flow. A shear layer of
$2\, R_\mathrm{j}$ in width is included in the initial rest mass density
and axial velocity profiles to keep numerical stability of the initial
jet. Elliptical and helical modes are induced at the inlet.

\noindent
Frequencies of the excited modes are computed from the observed
wavelengths of modes, $\lambda^\mathrm{obs}$, corrected for projection
effects, relativistic bulk motion and
wave pattern speed, $v_w=0.23\,c$, according to
$\omega=2\,\pi\,v_w/\lambda^{theor}$, where
\begin{equation}\label{lamb}
\lambda^{theor}=\frac{\lambda^\mathrm{obs}(1-v_w/c\,\cos\theta)}{\sin\theta},
\end{equation}
Three wavelengths have been identified in the simulation: a helical
$\lambda^\mathrm{sim}_1=4\,R_\mathrm{j}$ perturbation, a helical
$\lambda^\mathrm{sim}_2=25\,R_\mathrm{j}$ perturbation and an
elliptical $\lambda^\mathrm{sim}_3=50\,R_\mathrm{j}$ perturbation. The
wave speeds $v_{w,2}\simeq 0.38\,c$ and $v_{w,3}\simeq 0.2\,c$ are
estimated for the $\lambda^\mathrm{sim}_2$ and
$\lambda^\mathrm{sim}_3$ modes, respectively. An upper limit for
$v_{w,1}$ is given by the flow speed ($0.88\,c$). With these wave
speeds, $\lambda^\mathrm{sim}_1$ would be observed as a 2.27 mas
structure, $\lambda^\mathrm{sim}_2$ as a 3.37 mas structure and
$\lambda^\mathrm{sim}_3$ as a 5.5 mas structure. These are similar to
the shorter-wavelength structures found in the observations (2.27 mas
vs 2 mas, 3.37 and 5.5 mas vs 4 mas). It remains to be studied why the
simulations do not reproduce readily the longer modes.  The longest
mode, with the wave speed given in \cite{lz01}, corresponds to a simulated
wavelength of $150\,R_\mathrm{j}$, which is difficult
to reproduce even in the grid of this simulation, in particular
when shorter harmonics grow fast and disrupt the flow. 
The disruptive evolution may be caused by a number of factors, including 
the absence of a stabilizing magnetic field in the simulations.

\subsection{Simulation 3C\,273-B}

In the second simulation, we try to
study the effect of precession and injection of fast components on the
observed structures in the jet. The precession frequency is given by
the observed 15 yr periodicity of the position angle variations in the
inner jet (\cite{abr06}). The frequency of ejections of components is
set by the reported 1 yr periodicity (\cite{kri00}). The duration of
each ejection is estimated to be 2 months, from the approximate
inspiralling time from an orbit at $\sim 6\,R_G$ around a $5.5\, 10^8
\,M_\odot$.  Velocity of the components is taken as constant, with the
mean value of those given in \cite{abr06}, i.e., Lorentz factor $W
\sim 5$. The components are treated as shells of diameter
$0.5\,R_\mathrm{j}$ ejected at the center of the transversal grid. The
numerical grid for this simulation covers $30\,R_\mathrm{j}$ (axial)
$\times$ $6\,R_\mathrm{j}$ $\times$ $6\,R_\mathrm{j}$ (transversal),
i.e., 24 pc $\times$ 4.8 pc $\times$ 4.8 pc. The resolution of the
grid is 16 cells/$R_\mathrm{j}$ in both transversal directions and 32
cells/$R_\mathrm{j}$ in the direction of the flow.

\noindent
We find two structures in the simulation: a pinching perturbation with
a wavelength of $0.4\,R_\mathrm{j}$ caused by the injection of
components, and a helical perturbation with a wavelength of
$4.0\,R_\mathrm{j}$ associated with the precession. The wave speed
associated with these structures is $\le 0.98\,c$. This upper limit
for the wave speed results in observed wavelengths (from
Eq.~\ref{lamb}) of 0.6 and 6 mas, much smaller than the 2-4 mas and 18
mas expected from the observations. This results further supports the
identification of the shorter modes with K-H
instability modes. We also find that the 
15 yr precession period cannot account for the observed longer, 
18-mas wavelength. Thus, either the longer-wavelength structure
is driven by a different mechanism, or the 15 yr periodicity in the
source is not associated with the precession.

%
% BibTeX users please use
% \bibliographystyle{}
% \bibliography{}

%
% Non-BibTeX users please follow the syntax
% the syntax of "referenc.tex" for your own citations
%\input{referenc}

%%%%%%%%%%%%%%%%%%%%%%%%%%%%%%%%%%%%%%%%%%%%%%%%%%%%%%%%%%%%%%%%%%%%%%  }

%%%%%%%%%%%%%%%%%%%%%%%%%%%%%%%%%%%%%%%%%%%%%%%%%%%%%%%%%%%%%%%%%%%%%%

\printindex
\end{document}